\documentclass[12pt]{article}
\usepackage{graphicx,amssymb,amsfonts,amsmath,epsfig,color}

\makeatletter
\DeclareSymbolFont{AMSb}{U}{msb}{m}{n}
\DeclareSymbolFontAlphabet{\mathbb}{AMSb}
\renewcommand{\section}{\@startsection{section}{1}{\z@}%
                                    {-7ex \@plus -1ex \@minus -.2ex}%
                                    {2.5ex \@plus.2ex}%
                                    {\normalfont\large\scshape\centering}}
\renewcommand{\subsection}{\@startsection{subsection}{2}{\z@}%
                                       {-5ex \@plus -1ex \@minus -.2ex}%
                                       {1.5ex \@plus.2ex}%
                                       {\normalfont\normalsize\scshape}}
\renewcommand{\subsubsection}{\@startsection{subsubsection}{3}{\z@}%
                                       {-5ex \@plus -1ex \@minus -.2ex}%
                                       {1.5ex \@plus.2ex}%
                                       {\normalfont\normalsize\scshape}}

\renewcommand\@seccntformat[1]{\ignorespaces\csname #1name\endcsname\space
                               \csname the#1\endcsname.\quad}   
\newdimen\captionmargin
\newdimen\captionindent
\newdimen\captionwidth
\newcommand{\captionfont}{\slshape}
\newcommand\@captionlabel[1]{\textsc{#1:}\space}
\long\def\@makecaption#1#2{%
  \vskip\abovecaptionskip
  \captionwidth\hsize
  \advance\captionwidth -2\captionmargin
  \sbox\@tempboxa{\@captionlabel{#1}\captionfont #2}%
  \ifdim \wd\@tempboxa >\captionwidth
    \ifdim\captionindent>\z@
      \advance\captionwidth -\captionindent
      \hskip\captionindent
    \fi
    \hskip\captionmargin
    \parbox[t]{\captionwidth}{\leavevmode\hskip-\captionindent
      \@captionlabel{#1}\captionfont #2}%
  \else
    \global \@minipagefalse
    \hb@xt@\hsize{\hfil\box\@tempboxa\hfil}%
  \fi
  \vskip\belowcaptionskip}
\def\eqnarray{%
   \stepcounter{equation}%
   \def\@currentlabel{\p@equation\theequation}%
   \global\@eqnswtrue
   \m@th
   \global\@eqcnt\z@
   \tabskip\@centering
   \let\\\@eqncr
   $$\everycr{}\halign to\displaywidth\bgroup
       \hskip\@centering$\displaystyle\tabskip\z@skip{##}$\@eqnsel
      &\global\@eqcnt\@ne$\;\hfil{##}$\hfil
      &\global\@eqcnt\tw@$\;\displaystyle{##}$\hfil\tabskip\@centering
      &\global\@eqcnt\thr@@ \hb@xt@\z@\bgroup\hss##\egroup
         \tabskip\z@skip
      \cr}
\ifcase \@ptsize
\or
\or
\fi
  \divide\@tempdima\baselineskip
  \@tempcnta=\@tempdima
\makeatother
\begin{document}

\renewcommand{\theequation}{\arabic{section}.\arabic{equation}}
\renewcommand{\thefigure}{\arabic{figure}}
\newcommand{\gapprox}{%
\mathrel{%
\setbox0=\hbox{$>$}\raise0.6ex\copy0\kern-\wd0\lower0.65ex\hbox{$\sim$}}}
\textwidth 165mm \textheight 220mm \topmargin 0pt \oddsidemargin 2mm
\def\ib{{\bar \imath}}
\def\jb{{\bar \jmath}}

\newcommand{\ft}[2]{{\textstyle\frac{#1}{#2}}}
\newcommand{\be}{\begin{equation}}
\newcommand{\ee}{\end{equation}}
\newcommand{\bea}{\begin{eqnarray}}
\newcommand{\eea}{\end{eqnarray}}
\newcommand{\Identity}{{1\!\rm l}}
\newcommand{\cx}{\overset{\circ}{x}_2}
\def\CN{$\mathcal{N}$}
\def\CH{$\mathcal{H}$}
\def\hg{\hat{g}}
\newcommand{\bref}[1]{(\ref{#1})}
\def\espai{\;\;\;\;\;\;}
\def\zespai{\;\;\;\;}
\def\avall{\vspace{0.5cm}}
\newtheorem{theorem}{Theorem}
\newtheorem{acknowledgement}{Acknowledgment}
\newtheorem{algorithm}{Algorithm}
\newtheorem{axiom}{Axiom}
\newtheorem{case}{Case}
\newtheorem{claim}{Claim}
\newtheorem{conclusion}{Conclusion}
\newtheorem{condition}{Condition}
\newtheorem{conjecture}{Conjecture}
\newtheorem{corollary}{Corollary}
\newtheorem{criterion}{Criterion}
\newtheorem{defi}{Definition}
\newtheorem{example}{Example}
\newtheorem{exercise}{Exercise}
\newtheorem{lemma}{Lemma}
\newtheorem{notation}{Notation}
\newtheorem{problem}{Problem}
\newtheorem{prop}{Proposition}
\newtheorem{rem}{{\it Remark}}
\newtheorem{solution}{Solution}
\newtheorem{summary}{Summary}
\numberwithin{equation}{section}
\newenvironment{pf}[1][Proof]{\noindent{\it {#1.}} }{\ \rule{0.5em}{0.5em}}
\newenvironment{ex}[1][Example]{\noindent{\it {#1.}}}

\thispagestyle{empty}


\begin{center}

{\LARGE\scshape Singularity free gravitational collapse in an effective dynamical quantum spacetime
\par}
\vskip15mm

\textsc{R. Torres\footnote{E-mail: ramon.torres-herrera@upc.edu} and F. Fayos\footnote{E-mail: f.fayos@upc.edu} }
\par\bigskip
{\em
Department of Applied Physics, UPC, Barcelona, Spain.}\\[.1cm]

\vspace{5mm}

\end{center}

\begin{abstract}
We model the gravitational collapse of heavy massive shells including its main quantum corrections.
Among these corrections, quantum improvements coming from Quantum Einstein Gravity are taken into account, which provides us with an effective quantum spacetime. Likewise, we consider dynamical Hawking radiation by modeling its back-reaction once the horizons have been generated. Our results point towards a picture of gravitational collapse in which the collapsing shell reaches a minimum non-zero radius (whose value depends on the shell initial conditions) with its mass only slightly reduced. Then, there is always a rebound after which most (or all) of the mass evaporates in the form of Hawking radiation. Since the mass never concentrates in a single point, no singularity appears.
\end{abstract}

\vskip10mm
\noindent KEYWORDS: Gravitational Collapse, Black Holes, Hawking radiation, Quantum Gravity.



\setcounter{equation}{0}

\section{Introduction}

It is expected that a large enough object would collapse classically until a horizon forms. Then Hawking radiation would appear and the mass of the object should be reduced. Less is known on the details of the later evolution. In fact, a complete investigation of the process would require a complete consistent theory of quantum gravity together with the calculational tools to achieve a description of the scenario. Since such apparatus is not currently available, for the moment one can only resort to the study of toy models in which the known main quantum contributions are taken into account. By this means, one can try to probe some of the features that one could expect from a full theory of quantum gravity.

In this letter we will work in this direction. In our toy model two main simplifications will be carried out. First, we assume the existence of a spherically symmetric spacetime $\mathcal M$ in which the collapse takes place. Second, we choose as our collapsing object a \textit{thin shell}. In other words, we assume that the spacetime is split in two different regions $\mathcal M =\mathcal M^+ \cup \mathcal M^-$ with a common spherically symmetric timelike boundary $\Sigma= \partial \mathcal M^+ \cap \partial \mathcal M^-$ corresponding to the thin shell.

This second simplification deserves some comments. Clearly it means that we would be able to probe gravitational collapse only whenever the approximation in which one can neglect the shell thickness remains valid. An investigation of the conditions under which this is possible was carried out in \cite{ACVV} (see also \cite{O&R}). The authors considered a shell composed of a number $N$ of (s-wave) scalar particles with mass $m$ bound together by gravitational interaction. The $N$ particles form a radially localized bound state corresponding to a finite thickness shell, whose mean position approximately follows a classical collapse. Moreover, during the collapse the average shell thickness $d$ decreases according to \cite{ACVV}
\[
d\sim \frac{\hbar}{m} \left(\frac{R^2}{G_0 (N-1)}\right)^{1/3},
\]
where $G_0$ is Newton's gravitational constant.
On the other hand,
the fluctuations associated with the quantum nature of matter become dominant for a radius of the shell of the order of the Compton wavelength of the constituent quanta. So that they are negligible as long as \cite{ACVV}
\[
R\gg \hbar/m.
\]
In other words, if we want to probe the last stages of the collapsing phase by using the thin-shell approximation we should use a shell composed of a high number $N$ of very heavy particles (large $m$'s) and, thus, we would be using a heavy massive shell. Only under these conditions the results obtained using the thin-shell approximation are likely to be similar to the results that one would obtain in a real collapsing situation
\footnote{On the contrary, a \emph{light} shell with $M\lesssim m_p$ would possess a markedly quantum nature \cite{Hajicek}\cite{CCMPRRSV}.}.

With regard to the shell exterior region $\mathcal M^+$, we will describe it with a portion of an \textit{improved Schwarzschild solution} with mass equal to the shell mass. Specifically, we choose for the exterior region an effective improved solution coming from Quantum Einstein Gravity that incorporates quantum corrections to the classical solution. It does so by taking into account the effect of virtual gravitons. I.e., just as in quantum electrodynamics the virtual pairs imply the existence of a screening effect leading to a \textit{running electric charge}, when one considers the existence of virtual gravitons one obtains an \emph{antiscreening} effect leading to a \emph{running gravitational constant}, which is used to get the improved Schwarzschild solution (\cite{B&RIS} and references therein). A summary of this effective solution will be carried out in section \ref{secISS}. On the other hand and following this approach, the shell massless interior region $\mathcal M^-$ will have to be described by a portion of Minkowski's spacetime (what is equivalent to a massless improved Schwarzschild solution).

Since the improved exterior solution possesses horizons, the tunneling of virtual particles through them is expected to produce Hawking radiation. Thus, in section \ref{secHR} we will also model the effect of the back-reaction to the radiation in the effective solution. Then, in section \ref{secShell} we will consider the matching of the interior and exterior solutions through the spherically symmetric thin shell $\Sigma$ by using Israel's formalism \cite{IsraelM}. This will provide us with the shell evolution equation which will allow us to analyze its different \textit{attractive} and \textit{repulsive} contributions. Finally in section \ref{secNI}  the numerical integration of the evolution equation will be carried out and the results will be interpreted.

\section{Exterior: Improved Schwarzschild solution}\label{secISS}

As explained in the introduction, in order to model a collapsing shell we should first establish the exterior to that shell ($\mathcal M^+$). In this work we want the exterior to incorporate the main quantum corrections to the classical solution. This can be done by using a \textit{renormalization group improved} Schwarzschild solution found by Bonanno and Reuter \cite{B&RIS} that can be written as
\begin{equation}\label{RGISch}
ds^2=-\left(1-\frac{2 G(R) M}{R}\right) dt_S^2+\left(1-\frac{2 G(R) M}{R}\right)^{-1} dR^2+ R^2 d\Omega^2.
\end{equation}
where
\begin{equation}
G(R)=\frac{G_0 R^3}{R^3+\tilde{\omega} G_0 (R+\gamma G_0 M)},\label{runningG}
\end{equation}
$G_0$ is Newton's universal gravitational constant, $M$ is the mass measured by an observer at infinity and $\tilde{\omega}$ and $\gamma$ are constants coming from the non-perturbative renormalization group theory and from an appropriate ``cutoff identification", respectively.
The qualitative properties of this solution are fairly insensitive to the precise value of $\gamma$.
In this way, in order to simplify the calculations, it is usual to choose $\gamma=0$ \cite{B&RIS}\cite{B&RIV}.
On the other hand, $\tilde \omega$ can be found by comparison with the standard perturbative quantization of Einstein's gravity (see \cite{Dono} and references therein). It can be deduced that its precise value is $\tilde \omega=167/30\pi$, but again the properties of the solution do not rely on its precise value as long as it is strictly positive.

\begin{figure}
\includegraphics[scale=.8]{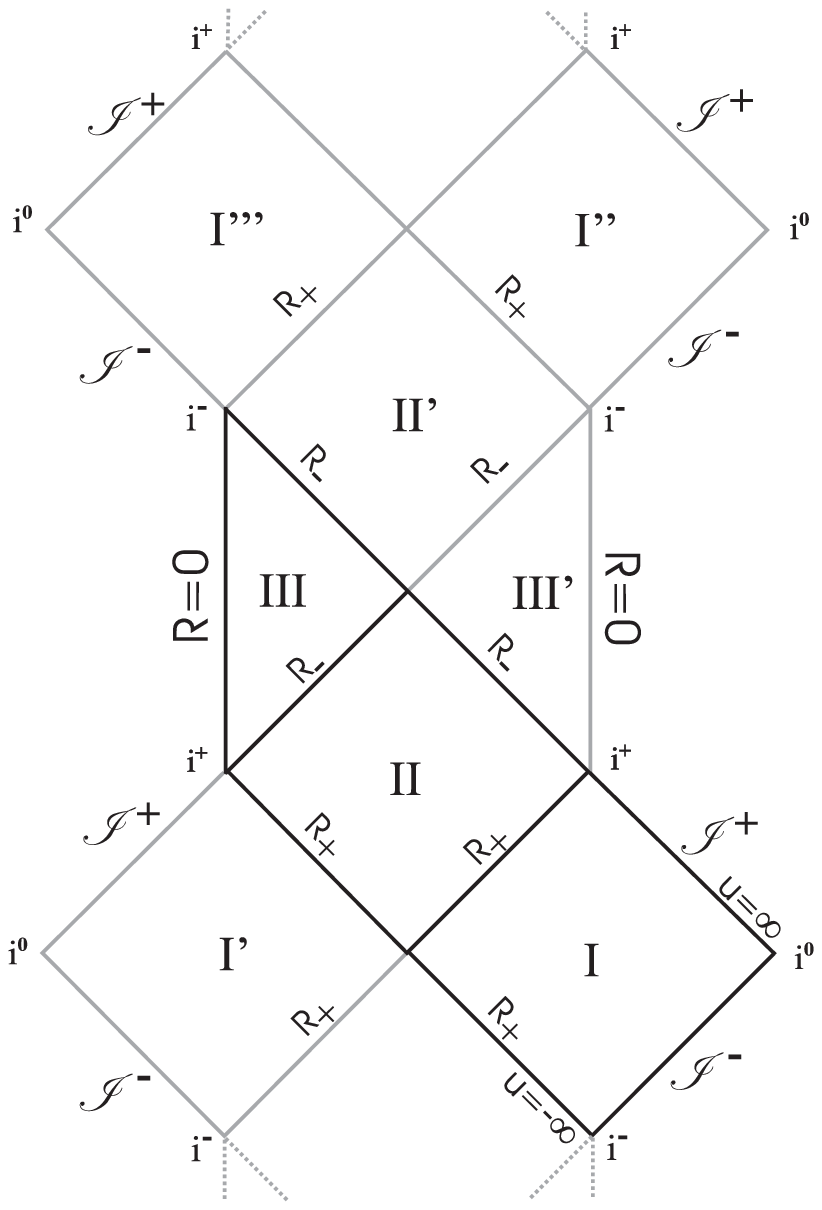}
\caption{\label{figstat} A Penrose diagram corresponding to the case $M>M_{cr}$. The regions drawn using a solid black line (I-II-III) correspond to the zone defined by the solution in Eddington-Finkelstein-like coordinates (\ref{ScEF}) with the null coordinate going from $u=-\infty$ to $u=\infty$. The regions drawn in grey correspond to extensions of this solution.}
\end{figure}

If we define
\[
f\equiv 1-\frac{2 G(R) M}{R},
\]
the horizons of the improved solution can be found by solving $f=0$.
Then, it is easy to see that the horizons correspond to the number of positive real solutions of a cubic equation and depend on the sign of its discriminant or, equivalently, on whether the mass is bigger, equal or smaller than a critical value $M_{cr}$.
In particular,
the value $\gamma=0$ implies
\[
M_{cr}=\sqrt{\frac{\tilde\omega}{G_0}}\simeq 1.33 m_p,
\]
where $m_p$ is the Planck mass.
If $M>M_{cr}$ then the equation $f=0$ has two positive real solutions $\{R_-,R_+\}$ satisfying $R_-<R_+$.
The existence of an inner solution $R_-$ represents a novelty with regard to the classical spacetime.
However, it is interesting to remark that it is a result common to different approaches to Quantum Gravity. (See, for example, \cite{B&RIS}\cite{Modes}\cite{Amel}\cite{Nicol}).
The outer solution $R_+$ can be considered as the \textit{improved Schwarzschild horizon}, i.e., the Schwarzschild horizon with quantum corrections taken into account. The `improvement' in this horizon can be made apparent for masses much bigger than Planck's mass if one expands $R_+$ in terms of $m_p/M$ obtaining
\[
R_+\simeq 2 G_0 M \left[1-\frac{(2+\gamma)}{8} \tilde\omega\ \left(\frac{m_p}{M}\right)^2\right].
\]
The global structure of the improved solution for $M>M_{cr}$ resembles the global structure of the Reissner-Nordstr\"{o}m spacetime with mass bigger than its charge ($M>|Q|$). A Penrose diagram corresponding to the improved solution for the $M>M_{cr}$ case is shown in fig. \ref{figstat}.
On the other hand, if $M=M_{cr}$ then there is only one positive real solution to the cubic equation and the global structure resembles that of a extremal Reissner-Nordstr\"{o}m solution ($M=|Q|$), whereas if $M<M_{cr}$ the equation has not positive real solutions.

\section{Hawking radiation from the horizons}\label{secHR}

We will now summarize the results on Hawking radiation in the quantum improved solution. A more complete description can be found in \cite{RPO}\cite{BHInt} which, in turn, are based on the tunneling approach by Parikh and Wilczek \cite{P&W}. We consider Hawking radiation coming out from an improved black hole satisfying $M>M_{cr}$ thanks to the tunneling process occurring both in the outer and in the inner horizons.

Since the coefficients of the metric do not depend on $t$ there is a killing vector $\partial/\partial t$ which is straightforwardly found to be timelike for $R>R_+$, lightlike for $R=R_+$, spacelike for $R_-<R<R_+$, lightlike for $R=R_-$ and timelike for $R<R_-$.
The possibility of tunneling is based on the fact that the killing vector is spacelike for $R_-<R<R_+$ (region II in fig.\ref{figstat}), what allows the existence of negative energy states.
Let us consider that a pair of photons is created in region II, where the 2-spheres are closed trapped surfaces.
A pair of test photons would be classically forced to move inwards until reaching $R=0$. However, for non-test photons energy conservation modifies this picture near the outer (inner) horizon since the positive energy photon produced in the pair could `tunnel' the outer (inner) horizon and move \emph{outwards} in region I (III ,respectively\footnote{Of course, one can also have the positive energy photon tunneling outwards in region I' (III', respectively).}).
This possibility, that would seem impossible in view of figure \ref{figstat}, is feasible because
energy conservation implies that, as the black hole mass would be reduced in such a process, the outer (inner) horizon would contract (expand, respectively) provoking the tunneling \cite{RPO}\cite{BHInt}.

In order to compute the tunneling rate we will rewrite the improved Schwarzschild's solution in Painlev\'e-like coordinates \cite{Pain} so as to have coordinates which are not singular at the horizons. In order to do this it suffices to introduce a new coordinate $t$ replacing the Schwarzschild-like time $t_S$ such that $t=t_S+h(R)$ and fix $h(R)$ by demanding the constant time slices to be flat. In this way one gets:
\begin{equation}
ds^2=-\left(1-\frac{2 G(R) M}{R}\right) dt^2+2 \sqrt{\frac{2 G(R) M}{R}} dt dR+ dR^2 + R^2 d\Omega^2,
\end{equation}
where $R$ can now take the values $0<R<\infty$.

Let us consider pair production occurring close to one of the horizons with the positive energy particle tunneling from region II outwards.
The standard results of the WKB method for the tunneling through a potential barrier that would be classically forbidden can be directly applied due to the infinite redshift near the horizon \cite{P&W}. In particular, the semiclassical emission rate will be given by $\Gamma \sim \exp\{-2 \mbox{Im} S\}$, where $S$ is the particle action. Therefore, we have to compute the imaginary part of the action for an
outgoing positive energy particle which crosses one of the horizons outwards from $R_{in}$ to $R_{out}$,
\begin{equation}\label{imS}
\mbox{Im} S=\mbox{Im} \int_{R_{in}}^{R_{out}} p_R dR= \mbox{Im} \int_{0}^{E} \int_{R_{in}}^{R_{out}} \frac{dR}{1-\sqrt{\frac{2 G(R;E')\cdot (M-E')}{R}}} (-dE').
\end{equation}
where we have used Hamilton's equation $\dot{R}=+dH/dp_R\rfloor_R$, the equation for null geodesics and the fact that the BH loses mass after the emission of a shell (i.e., $H=M-E'$) and, thus, $G(R)$ becomes $G(R;E')$, which stands for $G(R)$ with $M$ replaced by $M-E'$ (see \cite{RPO}\cite{BHInt} for details).

If we define
\begin{displaymath}
f(R;E')\equiv 1-\frac{2 G(R;E')\cdot (M-E')}{R}
\end{displaymath}
and
\[
g(R_\pm;E')\equiv
\frac{\partial f(R;E')}{\partial R}\rfloor_{R=R_\pm(E')},
\]
where $R_\pm(E')$ is the position of the outer (`+') or inner (`-') horizons  when $M$ is replaced by $M-E'$,
then,
by deforming the contour of integration so as to ensure that positive energy solutions decay in time,
one can then write (\ref{imS}) as
\begin{equation}\label{partchan}
\mbox{Im} S_\pm =\pm \int_{0}^{E}  \frac{2 \pi}{g(R_\pm;E')} dE'\ ,
\end{equation}
where the subindex `$+$' or `$-$' corresponds to the tunneling through the outer or inner horizon, respectively.

The semiclassical rate through every horizon will be
\begin{equation}\label{emprob}
\Gamma_\pm \sim e^{- 2 \mbox{\scriptsize Im} S_\pm }=\exp\left(\mp 4\pi \int_{0}^{E}  \frac{dE'}{g(R_\pm;E')} \right).
\end{equation}

When quadratic terms are neglected
we can develop Im $S$ up to first order in $E$ as
\[
\mbox{Im} S_\pm \simeq \mp\frac{2 \pi}{g(R_\pm,0)} E
\]
obtaining a thermal radiation for the quantum black hole ($\Gamma \sim \exp\{-E/T_\pm\}$) with (positive) temperature at every horizon
\begin{equation}\label{TQBH}
T_\pm=\pm \frac{g(R_\pm,0)}{4 \pi}=\pm \frac{1}{4 \pi}\left. \frac{\partial f}{\partial R}\right\rfloor_{R=R_\pm}.
\end{equation}

On the other hand, if we consider the full consequences of energy conservation, the distribution function for the emission of photons
can be written as (see \cite{K&K} --correcting the result of \cite{K&W}--)
\begin{displaymath}
<n(E)>=\frac{1}{\exp \left(2 \mbox{Im} S \right)-1}.
\end{displaymath}
What for our quantum corrected solution becomes, at every horizon,
\begin{equation}\label{nE}
<n(E)>_\pm =\frac{1}{\exp \left(\pm 4\pi \int_{0}^{E}  \frac{dE'}{g(R_\pm;E')} \right)
-1},
\end{equation}
with the additional requirement that, according to the properties of $g(R_\pm;E')$ (see \cite{RPO}\cite{BHInt}),
the energy of the emitted particles must satisfy $E\leq M-M_{cr}$.

Using (\ref{nE}), the flow of positive energy
due to the tunneled particles at the horizons can be written approximately as \cite{B&D}\cite{FN-S}
\begin{eqnarray}
L_\pm(M)\simeq \frac{1}{2 \pi}\int_0^{M-M_{cr}} <n(E)>_\pm \mathcal G_\pm E dE \nonumber\\
=\frac{1}{2 \pi}\int_0^{M-M_{cr}} \frac{E \mathcal{G}_\pm}{\exp\left(\pm 4\pi \int_{0}^{E}  \frac{dE'}{g(R_\pm;E')} \right)-1} dE,\label{lumi}
\end{eqnarray}
where
we have taken into account the possible backscattered region in the grey-body factor $\mathcal G_\pm$ (more details in \cite{RPO}\cite{BHInt}) and we are compelled to take into account in the integration limits that
the maximum energy of a radiated particle could be
$M-M_{cr}$.

\subsection{Modeling the back-reaction}

In order to modelize the evaporation, let us first write the improved Schwarzschild's metric (\ref{RGISch}) in terms of ingoing Eddington-Finkelstein-like coordinates $\{u,R,\theta,\varphi\}$, where
\[
u=t_S+\int^R \frac{dR'}{1-2 G(R') M/R'}\ ,
\]
as
\begin{equation}\label{ScEF}
ds^2=-\left(1-\frac{2 G(R) M}{R}\right) du^2+2 du dR + R^2 d\Omega^2.
\end{equation}
This solution does not reflect the back-reaction associated to the lost of mass due to the tunneling effect. However, we can model the mass lost taking into account that, whenever a pair of virtual particles is created, when the particle with positive energy escapes to infinity its companion, with negative energy, falls inwards and reduces the total mass. In this way, if we consider negative energy massless particles following ingoing null geodesics $u=$constant, the mass becomes a decreasing function $M(u)$. The metric which incorporates the effect of the decreasing mass due to the ingoing null radiation
is (\ref{ScEF}) with $M$ replaced by $M(u)$, i.e., it corresponds to an \textit{improved} ingoing Vaidya solution \cite{B&RIV} that for the $\gamma=0$ case that we will treat in our specific model takes the form
\begin{equation}\label{Vaid}
ds^2=-\left(1-\frac{2 \tilde G(R) M(u)}{R}\right) du^2+2 du dR + R^2 d\Omega^2,
\end{equation}
where, from (\ref{runningG}), the \textit{running} gravitational constant takes the form
\begin{equation}
\tilde{G}(R)=\frac{G_0 R^2}{R^2+\tilde{\omega} G_0}. \label{GR}
\end{equation}
Let us comment that when one considers the back-reaction the solutions to $f(u,R)=1-2 \tilde G(R) M(u)/R=0$ become dynamical (marginally trapped) horizons $R_\pm (u)$.
On the other hand, the flux of negative energy particles directed inwards equals the flux of outgoing radiated particles and, therefore\footnote{It is important to remark that in this section we are dealing only with the improved Schwarzschild solution. In the complete collapsing model, Hawking radiation will be directed towards the shell and the expression equivalent to (\ref{difM}) will have to be reconsidered. See section \ref{secNI}.},
\begin{equation}\label{difM}
\frac{d M(u)}{du}=- L_{Total}(M(u)),
\end{equation}
where $L_{Total}$ comes from the combination of fluxes $L_+$ and $L_-$ coming from the outer and inner horizons, respectively.

\section{Collapsing model}\label{secShell}

In order to model the collapsing thin shell we will now use Israel's formalism \cite{IsraelM}\cite{Poisson}. We, therefore, assume that the spherically symmetric spacetime $\mathcal M$ is split in two different regions $\mathcal M =\mathcal M^+ \cup \mathcal M^-$ with a common spherically symmetric timelike boundary $\Sigma= \partial \mathcal M^+ \cap \partial \mathcal M^-$: The thin shell.
We choose that the shell is described by coordinates $\{y^a\}=\{\tau, \theta_\Sigma, \varphi_\Sigma\}$ such that $\tau$ is the shell proper time and the parametric equations from $\mathcal M^+$ can be locally written in Eddington-Finkelstein-like coordinates as $\{{x^+}^\mu(y^a)\}=\{u(\tau),R(\tau),\theta=\theta_\Sigma,\varphi=\varphi_\Sigma\}$ (and similarly for $\mathcal M^-$).
In order to have a well-defined geometry at $\mathcal M$ the first fundamental forms (or induced line elements) of the boundary $\Sigma$
\begin{displaymath}
h_{ab}^\pm\equiv g_{\mu\nu}^\pm {e^\pm}^\mu_a {e^\pm}^\nu_b,
\end{displaymath}
where ${e^\pm}^\alpha_a=\partial {x^\pm}^\alpha/\partial y^a$,
must agree when computed from $\mathcal M^+$ or $\mathcal M^-$, i.e., $h_{ab}^+=h_{ab}^-$.
A first important consequence of this fact in the case of spherically symmetric scenarios  is that the \textit{areal coordinates} ($R$) for the interior and exterior regions must agree on the shell. In this way, if we want to describe the evolution of the shell, the same function $R(\tau)$ can be used from either the point of view of $\mathcal M^+$ or $\mathcal M^-$.

Let $\mathbf n$ be the unit spacelike vector ($\mathbf n \cdot \mathbf n=1$) pointing from $\mathcal M^-$ to $\mathcal M^+$. If we now define $\zeta$ as a coordinate such that $\mathbf n=\partial/\partial \zeta$ with $\zeta=0$ at the hypersurface $\Sigma$, the energy-momentum tensor of the spacetime would have the form
\begin{equation}
T_{\mu\nu}=S_{\mu\nu} \delta(\zeta) + T_{\mu\nu}^+ \theta(\zeta)+  T_{\mu\nu}^- \theta(-\zeta),
\end{equation}
where $\theta(x)$ is the Heaviside step-function and $S_{\mu\nu}$, which is tangent to $\Sigma$, is the energy-momentum tensor of the hypersurface.
The extrinsic curvature or second fundamental form of $\Sigma$ is defined as
\begin{displaymath}
{K^\pm}_{ab}\equiv {n^\pm}_{\mu;\nu} {e^\pm}^\mu_a {e^\pm}^\nu_b
\end{displaymath}
and it is related to the energy-momentum tensor of the hypersurface through the Lanczos equations
\begin{equation}
[K_{ab}]=8\pi G_0 \left(S_{ab}-\frac{1}{2} h_{ab} S\right),\label{Lanczos}
\end{equation}
where
$[K_{ab}]=K_{ab}^+-K_{ab}^-$ and we are using the shell energy-momentum three-tensor $S_{ab}=S_{\alpha\beta} {e}^\alpha_a {e}^\beta_b$ and $S\equiv S_{ab} h^{ab}$.

In our model we will consider that $\Sigma$ is composed of radially moving non-interacting particles, so that
\begin{equation}
S_{ab}=\sigma v_a v_b,
\end{equation}
where $\sigma$ is the mass-energy density of the layer and $\mathbf v =d/d\tau$ is the shell 4-velocity.
As seen from the exterior region $\mathcal M^+$ we will have
\begin{equation}
\mathbf{v}^+=\dot{u}\frac{\partial}{\partial u}+\dot{R}\frac{\partial}{\partial R}\ \ \ , \ \ \ \mathbf{n}^+=-\dot{R} du + \dot{u} dR,
\end{equation}
where the dot stands for derivative with respect to $\tau$.
On the other hand, using the metric (\ref{Vaid}), the normalization condition $\mathbf v^+ \cdot \mathbf v^+=-1$ implies
\begin{equation}
Y(\tau)=\frac{\dot{R}+\beta^+}{f},\label{up}
\end{equation}
where $Y(\tau)\equiv\dot{u}$ and $\beta^+\equiv \sqrt{\dot{R}^2+f}$. From this last definition we see that a necessary condition for $\Sigma$ to be timelike is
\begin{equation}
\dot{R}^2+f\geq 0.\label{minspeed}
\end{equation}
In physical terms this can be interpreted as requiring the collapsing shell to have a minimum speed while it traverses the region where the 2-spheres are closed trapped surfaces ($f<0$).
On the other hand, the extrinsic curvature of $\Sigma$ when computed from the exterior region $\mathcal{M}^+$ using (\ref{Vaid}) is
\begin{eqnarray}
K_{\tau\tau}^+&=&\frac{1}{\beta^+}\left(\ddot{R}+\frac{\tilde G(R) M}{R^2}+ \frac{\tilde G(R) Y}{R} \frac{dM}{d\tau}-\frac{M}{R} \frac{d\tilde G}{dR}\right)\label{K+11}\\
K_{\theta\theta}^+&=&- R(\tau) \beta^+\label{K+22}\\
K_{\varphi\varphi}^+&=& \sin^2\theta K_{\theta\theta}^+.
\end{eqnarray}

Since the interior of the shell has no mass, one can get the results for the interior region $\mathcal M^-$ from the previous ones with $M=0$. In particular, (\ref{Vaid}) becomes Minkowski's solution and
\begin{eqnarray}
\beta^-&\equiv& \sqrt{\dot{R}^2+1}\\
K_{\tau\tau}^-&=&\frac{1}{\beta^-} \ddot{R} \label{K-11}\\
K_{\theta\theta}^-&=&- R(\tau) \beta^-\label{K-22}\\
K_{\varphi\varphi}^-&=& \sin^2\theta K_{\theta\theta}^-.
\end{eqnarray}

\subsection{The evolution of the shell}

The Lanczos equations (\ref{Lanczos}) provide us with
\begin{eqnarray}
[K_{\tau\tau}]&=& 4 \pi \sigma\\
\left[K_{\theta\theta}\right]&=& 4 \pi \sigma R^2
\end{eqnarray}
which can be easily combined using (\ref{K+11}, \ref{K+22}, \ref{K-11}, \ref{K-22}) in order to obtain
an equation for the mass-energy density
\begin{equation}
\sigma = \frac{\beta^- -\beta^+}{4 \pi R}
\end{equation}
and an \textit{evolution equation}
\begin{equation}
\ddot{R} = \frac{\beta^-}{2 R}(\beta^+-\beta^-)+\frac{\beta^-}{\beta^+-\beta^-}\left(
\frac{\tilde G Y}{R} \frac{dM}{d\tau}-\frac{M}{R} \frac{d\tilde G}{dR}\right). \label{evoleq}
\end{equation}

In order to analyze the meaning of this equation, first note that $f \leq 1$, $\beta^-\geq 1$ and that $\beta^+\leq\beta^-$. Then we can differentiate three \emph{force terms} in the right hand side of the equation. First, the usual term coming from the self-gravitating shell
\[
\mathcal{F}_S \equiv \frac{\beta^-}{2 R}(\beta^+-\beta^-) ,
\]
that satisfies $\mathcal F_S \leq 0$ which implies that the shell, in the regions where the quantum corrections are small, is accelerated inwards. Second, the Hawking-radiation term
\[
\mathcal F_H \equiv \frac{\beta^-}{\beta^+-\beta^-} \frac{\tilde G Y}{R} \frac{dM}{d\tau}
\]
that satisfies $\mathcal F_H \geq 0$. Specifically, in this approach this term does not contribute previous to the formation of a horizon since there is not Hawking radiation. However, once a horizon appears one has $dM/d\tau<0$ due to the negative energy coming from the horizon which is absorbed by the shell.
Clearly, $\mathcal F_H \geq 0$ implies that the absorbtion of negative energy contributes with a force opposed to the collapse. Finally, the third term,
\begin{equation}
\mathcal F_Q \equiv \frac{\beta^-}{\beta^- -\beta^+} \frac{M}{R} \frac{d\tilde G}{dR} \label{fQ}
\end{equation}
satisfies $\mathcal F_Q \geq 0$ since $d\tilde{G}/dR \geq 0$. The term can be interpreted as a force of quantum origin due to the antiscreening effect of virtual gravitons. This force is opposed to the collapse, however, it is easy to see that its effect is only relevant for $R\lesssim 10 l_p$.

\section{Results from the numerical integration}\label{secNI}

In order to get the evolution of the shell we should numerically integrate (\ref{evoleq}) using that, from (\ref{up}),
\begin{equation}\label{MassDif}
\frac{dM}{d\tau}=Y \ \frac{dM}{du}.
\end{equation}
We should also take into account that $dM/du$ is a piecewise defined function satisfying
\begin{itemize}
\item $dM/du=0$ if the areal radius $R(\tau)$ of the shell satisfies $R(\tau)\geq R_+(\tau)$. I.e., according to the tunneling picture, when the collapse has not still reached the stage in which a dynamical outer horizon appears, no Hawking radiation exists.
\item $dM/du=- \alpha L_+$ if $R_-(\tau)< R(\tau)< R_+(\tau)$, where we should use (\ref{lumi}) in order to take into account the radiation coming from the horizon $R_+$ which is absorbed by the shell. I.e., once the collapse of the shell generates an outer horizon, negative energy Hawking radiation diminishes the shell mass. The absorption coefficient $\alpha$ informs us about the fraction of energy that is absorbed by the shell. In our model we will consider $\alpha \simeq 1$.
\item $dM/du=- \alpha (L_+ + L_-)$ if $R(\tau) \leq R_-(\tau) $, where we should use (\ref{lumi}) in order to take into account the radiation coming from both the horizons $R_+(\tau)$ and $R_-(\tau)$. I.e. when the collapse of the shell also generates an inner horizon then the negative energy coming from both the outer and the inner horizon contribute to the mass loss of the shell.
\end{itemize}

\begin{figure}
\includegraphics[scale=1]{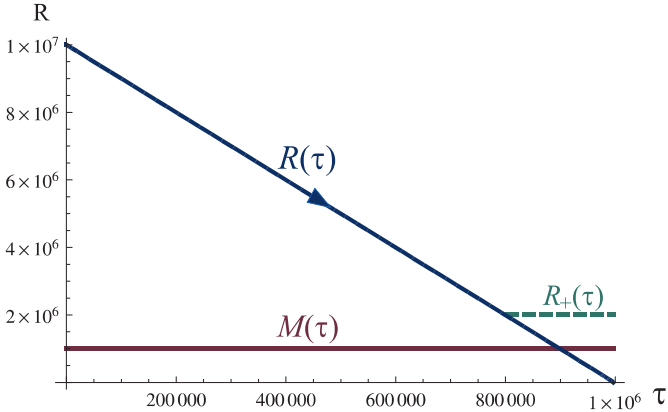}
\caption{\label{evolini} Results of the numerical computation of the evolution equation for a collapsing massive shell, where the areal radius and the mass of the shell are plotted as functions of the proper time of the shell. The figures specifically exemplifies the case $M(\tau=0)=10^6,\ R(\tau=0)=10^7 (>R_+),\ R'(\tau=0)=-10$, although the results are generic. First, there are no horizons and the shell relentlessly collapses with constant mass. Then the mass is concentrated enough to generate an outer horizon (at $\tau\simeq 799595$, for the chosen initial values) what activates Hawking radiation. However, the mass reduction is negligible in the period in which $R(\tau)$ decreases. (Note that the generation of an inner horizon and later behaviour of the shell are not shown in this figure).}
\end{figure}

The numerical integration of the system of differential equations composed of the evolution equation (\ref{evoleq}) and the mass equation (\ref{MassDif}) during the collapse of the shell from its initial areal radius $R(\tau=0)> R_+(\tau=0)$ until $R(\tau)\sim R_-(\tau)$ is exemplified in fig. \ref{evolini}. In this figure we can see how the shell collapses creating an outer horizon $R_+(\tau)$. However, the figure does not show neither the generation of the inner horizon nor the shell's later behaviour since they are indistinguishable at the chosen drawing scale. In the same figure we have also plotted the evolution of the mass. It is important to note that, even if negative energy Hawking radiation is being absorbed once the shell generates its outer horizon, the initial mass is so big and the collapse so fast that the mass remains practically constant at this stage. In this way, one can conclude that the behaviour of the heavy massive shell previous to the generation of the \emph{inner} horizon is practically identical to the behaviour of a collapsing classical shell.

\begin{figure}
\includegraphics[scale=1]{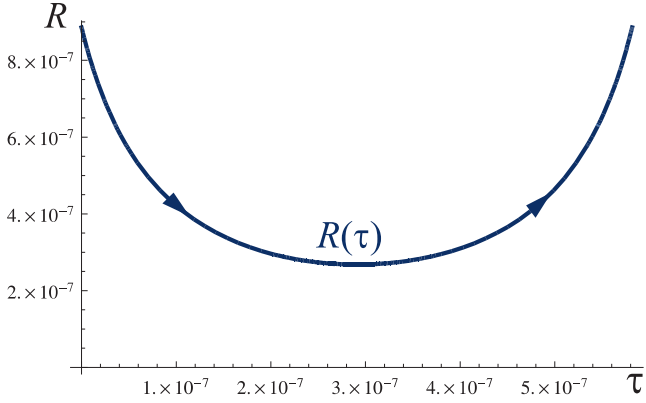}
\caption{\label{rebound} A continuation of fig. \ref{evolini} in the BH interior once an inner horizon has been generated and in which the rebound is explicitly shown. The shell $R(\tau)$ crosses the inner horizon at $\tau=0$. Specifically, $R_-(\tau=0)\simeq 8.860\cdot 10^{-7}$, for the particular initial conditions chosen. The shell rebounds at $\tau \simeq 2.908\cdot 10^{-7}$. Then, it approaches the inner horizon $(u\simeq \infty, R\simeq R_-)$ at $\tau \simeq 5.81672\cdot 10^{-7}$. Strictly speaking, the trajectory described in the figure should not be interpreted literally, but simply as a strong indication of the possible existence of a rebound. Although we exemplify this with a specific case, the behaviour is generic for heavy massive shells due to the divergent character of the repulsive term $\mathcal F_Q$ (\ref{fQ}) at $R=0$.}
\end{figure}

A continuation of fig. \ref{evolini} is shown in figure \ref{rebound}. It describes the situation when the antiscreening effect of the virtual gravitons have already forced the generation of an inner horizon $R_-$. We specifically show the evolution of the shell beyond this horizon. As has been exemplified in the graphic, the numerical integration of the system of differential equations provides us with a generic rebound of the shell which can be interpreted as due to the repulsive antiscreening force $\mathcal F_Q$ (\ref{fQ}).

After the rebound takes place the shell again approaches $R_-(\tau)$, what implies that $f$ tends to zero. Then, since $\dot R >0$, equation (\ref{up}) tells us that $Y(\tau)=\dot u$ will diverge and that the shell radius will reach the inner horizon $(u=\infty, R=R_-)$ (see fig. \ref{figstat}).

If we just consider the results obtained so far\footnote{I.e., here we are not taken into account possible instabilities and quantum gravity effects that will be commented later.}, it is known \cite{RPO}\cite{BHInt} that, from the point of view of an exterior observer living in region $I$ she would see Hawking radiation coming from the surroundings of (only) the outer horizon and she would notice an increase in the perceived flux varying from an initial negligible amount until reaching a maximum for $M\gtrsim M_{cr}$. After this maximum the flux tends towards zero while $M$ reaches the value $M_{cr}$.

\begin{figure}
\includegraphics[scale=1]{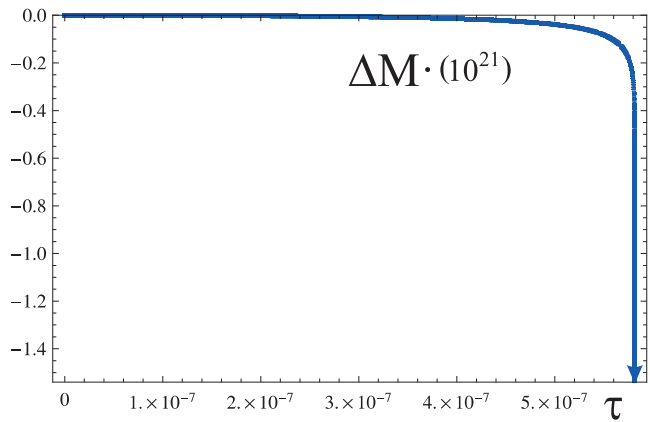}
\caption{\label{MassInc} The decrease of the shell mass, according to the shell, as it evolves in region III. Specifically, we show $\Delta M(\tau)\equiv M(\tau)-M(\tau=0)$. In order to plot this figure a numerical computation has been carried out using the same initial conditions as in fig. \ref{rebound}. In particular, $M(\tau=0)\simeq 10^6$.
Note that the rebound ($\tau \simeq 2.908\cdot 10^{-7}$) happens without any substantial change in the shell mass. Only when the shell is close enough to the value $\tau \simeq 5.816724\cdot 10^{-7}$ in which it reaches (for the chosen initial conditions) the inner horizon $(u=\infty,R=R_-)$ the decrease in mass due to Hawking radiation becomes huge. Note that, previous to that, the decrease in the mass function was so small that we have amplified $\Delta M$ by a factor of $10^{21}$ in order to the decrease to be noticeable.}
\end{figure}

From the point of view of the shell things look quite different. We have seen that the shell collapses with negligible mass decrease while $\dot R<0$. However, in its short travel in region III the shell reaches the inner horizon  $(u=\infty, R=R_-)$ after the rebound while managing to reduce its mass from a large value down to the Planckian value $M_{cr}$ (in agreement with the external observer). It is not hard to understand this fast decrease from the point of view of the shell since we have $\dot M=Y dM/du$ with $Y(\tau)$ diverging at $(u=\infty, R=R_-)$.
In figure \ref{MassInc} we show the results of the mass evolution as a function of the shell proper time.
There one can visualize how it is only
when the shell is very close to the inner horizon $(u=\infty, R=R_-)$ that all the negative energy from the horizons is received in the form of an sudden implosion of negative energy radiation that causes the shell to lose most of its mass.

\begin{figure}
\includegraphics[scale=1]{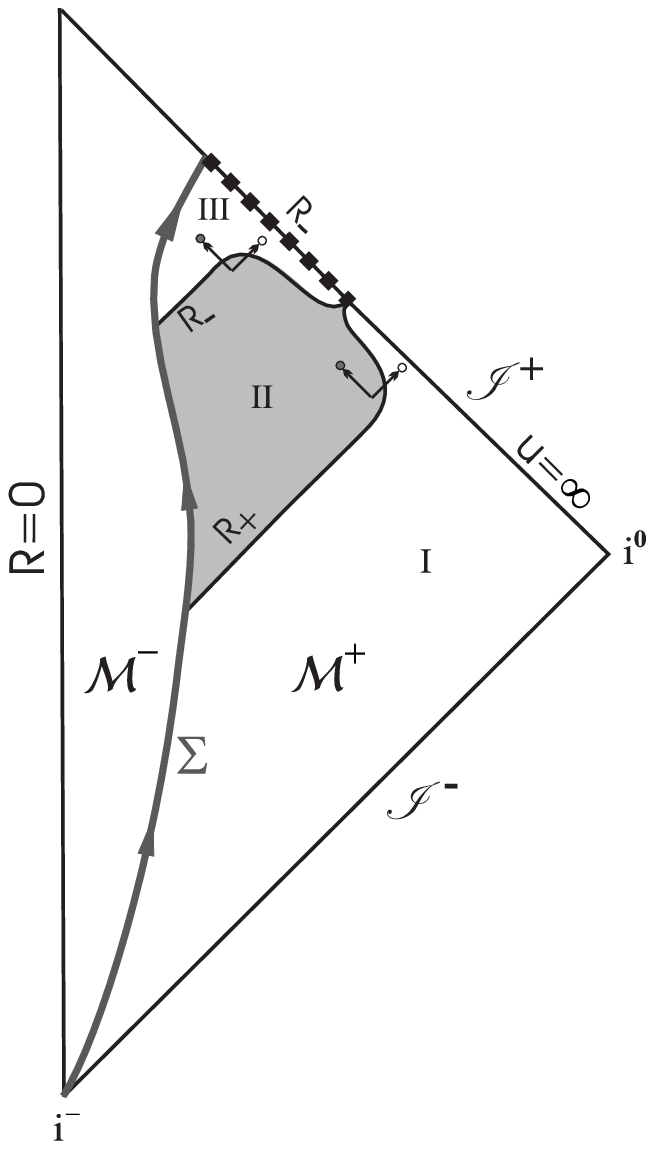}
\caption{\label{collapse} A Penrose diagram of a collapsing massive shell according to our results. The exterior advanced time is in the range $-\infty\leq u <\infty$ with the shell originating in the timelike past infinity $i^-$. The backreaction to the tunneling of particles in the inner and outer horizon is reflected in that the inner horizon expands while the outer horizon shrinks. The tunneling of particles generated in region II through the outer and inner horizon has been schematically shown: A pair is created in region II. The negative energy particle (darker circle) falls towards $R=0$ while the positive energy particle (lighter circle) \textit{tunnels} outwards and then follows the outgoing direction. On the other hand, the dashed line on the inner horizon $(u=\infty,R=R_-)$ represents our ignorance on the resolution of the black hole endogenous instability.}
\end{figure}

By collecting our results we have drawn the complete picture of the gravitational collapse of the shell in a Penrose diagram (fig. \ref{collapse}). Since the improved solution has an \textit{endogenous instability} at $(u=\infty, R=R_-)$ \cite{BHInt} the actual behaviour of the spacetime around and beyond this horizon is not clear. This has been represented by a dashed line in the figure.

\section{Conclusions}

In this letter we have described the collapse of a heavy massive thin shell. We have commented that only for large enough masses the behaviour is expected to be similar to the strictly classical one during the collapsing phase. In this way, the generation of an (outer) horizon $R_+$ will be compulsory. We have seen that, once the horizon has been generated, a negligible amount of Hawking radiation will leave the horizon surroundings towards the future null infinity (in the form of positive energy particles --mostly photons) and towards the collapsing shell (in the form of negative energy particles --again, mostly photons) during this phase. There are two reasons why Hawking radiation is negligible at this stage. First, the numerical computations show that the time taken for the shell between the generation of the outer and the inner horizon is (and should be (\ref{minspeed})) small. Second, the initial mass of the modeled shell is large and, therefore, Hawking radiation from the outer horizon (that approximately behaves as $- dM/du \sim 1/M^2$ at this stage) is negligible.

We have seen that there is a repulsive force $\mathcal F_Q$ acting on the shell due to the antiscreening effect of the virtual gravitons. However, this force only becomes relevant for $R\lesssim 10 l_p$, i.e., its effects appear while the shell is gravitationally trapped inside the outer horizon ($f\rfloor_\Sigma<0$).
Therefore, from the dynamical point of view modeled by the collapsing shell, it can be interpreted that this quantum repulsive force is the ultimate responsible that allows the shell to stop being gravitationally trapped or, equivalently, that causes the later $f\rfloor_\Sigma>0$ behaviour after generating an inner horizon.

Moreover, since $(\beta_+ - \beta_- )\sim R$ and $G'(R) \sim R$ for $R\sim 0$, the repulsive force (\ref{fQ}) eventually creates an \textit{impassable barrier} when $R\sim 0$ due to the divergence of $\mathcal F_Q \sim 1/R$. In this way, there will always be a rebound point for the shell underneath the inner horizon $R_-(\tau)$. This rebound has been explicitly displayed with the help of some numerical computations in fig. \ref{rebound}. Therefore, our results point towards the conclusion that the total collapse of the shell and the subsequent creation of a singularity could be avoided.

Once the rebound point has been reached, the evolution continues with an expanding areal radius that approaches the inner horizon $R_-(\tau)$, but this time with a diverging exterior null time $u$. We have seen that, from the point of view of the shell, this means that it is only when it is very close to this horizon that it suddenly receives a huge amount of negative energy coming from the outer and inner horizons, what would be interpreted by the shell as an implosion of negative energy radiation. Due to this fact, the shell should lose most (or all) of its mass almost instantaneously according to its own proper time. Strictly speaking, the model indicates that the mass should suddenly reach the Planckian value $M_{cr}$. However, this cannot be guaranteed since the exterior region has an endogenous instability \cite{BHInt} at the inner horizon ($u=\infty, R=R_-$) whose resolution is not at all clear.
Thus, the ultimate result of the collapse (remnant, total evaporation, etc.) remains unclear. In our opinion, the resolution of the inner horizon instability and the knowledge of the ultimate result of the collapse would probably require the help of a full Theory of Quantum Gravity.

\end{document}